# Independence of Maxwell's equations: A Bäcklund-transformation view

**C. J. Papachristou** [(a)] **and A. N. Magoulas** [(b)]

[(a)] Department of Physical Sciences, Hellenic Naval Academy
[(b)] Department of Electrical Engineering, Hellenic Naval Academy

**Abstract.** It is now widely accepted that the Maxwell equations of Electrodynamics constitute a self-consistent set of four independent partial differential equations. According to a certain school of thought, however, half of these equations – namely, those expressing the two Gauss' laws for the electric and the magnetic field – are redundant since they can be "derived" from the remaining two laws and the principle of conservation of charge. The status of the latter principle is thus elevated to a law of Nature more fundamental than, say, Coulomb's law. In this note we examine this line of reasoning and we propose an approach according to which the Maxwell equations may be viewed as a Bäcklund transformation relating fields and sources. The conservation of charge and the electromagnetic wave equations then simply express the integrability conditions of this transformation.

**Keywords:** Classical electrodynamics, Maxwell's equations, Bäcklund transformations

## 1. Is Gauss' law of Electrodynamics redundant?

As we know, the *Maxwell equations* describe the behavior (that is, the laws of change in space and time) of the electromagnetic (e/m) field. This field is represented by the pair $(\vec{E},\vec{B})$, where $\vec{E}$ and $\vec{B}$ are the electric and the magnetic field, respectively. The Maxwell equations additionally impose certain boundary conditions at the interface of two different media, while certain other physical demands are obvious (for example, the e/m field must vanish away from its localized "sources", unless these sources emit e/m radiation).

The Maxwell equations are a system of four partial differential equations (PDEs) that is self-consistent, in the sense that these equations are compatible with one another. The self-consistency of the system also implies the satisfaction of two important conditions that are physically meaningful:

- the *equation of continuity*, related to *conservation of charge*; and
- the *e/m wave equation* in its various forms.

We stress that these conditions are *necessary but not sufficient* for the validity of the Maxwell system. Thus, although every solution $(\vec{E},\vec{B})$ of this system obeys a wave equation separately for the electric and the magnetic field, an arbitrary pair of fields $(\vec{E},\vec{B})$, each field satisfying the corresponding wave equation, does not necessarily satisfy the Maxwell system itself. Also, the principle of conservation of charge *cannot replace* any one of Maxwell's equations. These remarks are justified by the fact that the aforementioned two necessary conditions are derived by *differentiating* the Maxwell system and, in this process, part of the information carried by this system is lost. [Recall, similarly, that cross-differentiation of the Cauchy-Riemann relations of complex analysis yields the Laplace equation (see Sec. 2) by which, however, we cannot recover the Cauchy-Riemann relations.]

The differential form of the Maxwell equations is

$$
\begin{aligned}
&(a)\quad \vec{\nabla}\cdot\vec{E}=\frac{\rho}{\varepsilon_0} &\qquad &(c)\quad \vec{\nabla}\times\vec{E}=-\frac{\partial\vec{B}}{\partial t}\\
&(b)\quad \vec{\nabla}\cdot\vec{B}=0 &\qquad &(d)\quad \vec{\nabla}\times\vec{B}=\mu_0\vec{J}+\varepsilon_0\mu_0\frac{\partial\vec{E}}{\partial t}
\end{aligned}
\tag{1}
$$

where $\rho$, $\vec{J}$ are the charge and current densities, respectively (the "sources" of the e/m field). Both the fields and the sources are functions of the spacetime variables $(x,y,z,t)$. Equations (1*a*) and (1*b*), which describe the *div* of the e/m field at any moment, constitute *Gauss' law* for the electric and the magnetic field, respectively. In terms of physical content, (1*a*) expresses the Coulomb law of electricity, while (1*b*) rules out the possibility of existence of magnetic poles analogous to electric charges. Equation (1*c*) expresses the *Faraday-Henry law* (law of e/m induction) and Eq. (1*d*) expresses the *Ampère-Maxwell law*. Equations (1*a*) and (1*d*), which contain the sources of the e/m field, constitute the *non-homogeneous* Maxwell equations, while Eqs. (1*b*) and (1*c*) are the *homogeneous* equations of the system.

By taking the *div* of (1*d*) and by using (1*a*), we obtain the *equation of continuity*, which physically expresses the *principle of conservation of charge* (see, e.g., [1], Sec. 9.6):

$$
\vec{\nabla}\cdot\vec{J}+\frac{\partial\rho}{\partial t}=0 \tag{2}
$$

Although the charge and current densities on the right-hand sides of (1*a*) and (1*d*) are chosen freely and are considered known from the outset, relation (2) places a severe restriction on the associated functions. A different kind of differentiation of the Maxwell system (1), by taking the *rot* of (*c*) and (*d*), leads to separate wave equations (or modified wave equations, depending on the medium) for the electric and the magnetic field (see, e.g., [1], Sec. 10.4).

In most textbooks on electromagnetism (e.g., [2–6] and many more) the Maxwell equations (1) are treated as a consistent set of four independent PDEs. A number of authors, however, have doubted the independence of this system. Specifically, they argue that (1*a*) and (1*b*) – the equations for the *div* of the e/m field, expressing Gauss' law for the corresponding fields – are redundant since they "may be derived" from (1*c*) and (1*d*) in combination with the equation of continuity (2). If this is true, Coulomb's law – the most important experimental law of electricity – loses its status as an independent law and is reduced to a derivative theorem. The same can be said with regard to the non-existence of magnetic poles in Nature.

As far as we know, the first who doubted the independent status of the two Gauss' laws in electrodynamics was Julius Adams Stratton in his 1941 famous (and, admittedly, very attractive) book [7]. His reasoning may be described as follows:

By taking the *div* of (1*c*), the left-hand side vanishes identically while on the right-hand side we may change the order of differentiation with respect to space and time variables. The result is:

$$\frac{\partial}{\partial t}\left(\vec{\nabla}\cdot\vec{B}\right)=0 \tag{3}$$

On the other hand, by taking the *div* of (1*d*) and by using the equation of continuity (2), we find that

$$\frac{\partial}{\partial t}\left(\vec{\nabla}\cdot\vec{E}-\frac{\rho}{\varepsilon_0}\right)=0 \tag{4}$$

And the line of argument continues as follows: According to (3) and (4), the quantities $\vec{\nabla}\cdot\vec{B}$ and ($\vec{\nabla}\cdot\vec{E}-\rho/\varepsilon_0$) are constant in time at every point $(x,y,z)$ of the region $\Omega$ of space that concerns us. *If* we now assume that there has been a period of time during which no e/m field existed in the region $\Omega$, then, in that period,

$$\vec{\nabla}\cdot\vec{B}=0 \quad \text{and} \quad \vec{\nabla}\cdot\vec{E}-\rho/\varepsilon_0=0 \tag{5}$$

identically. Later on, although an e/m field did appear in $\Omega$, the left-hand sides in (5) continued to vanish everywhere within this region since, as we said above, those quantities are time constant at every point of $\Omega$. Thus, by the equations for the *rot* of the e/m field and by the principle of conservation of charge – the status of which was elevated from derivative theorem to fundamental law of the theory – we derived Eqs. (5), which are precisely the first two Maxwell equations (1*a*) and (1*b*)!

According to this reasoning, the electromagnetic theory is not based on four independent Maxwell equations but rather on *three* independent equations only; namely, the Faraday-Henry law (1*c*), the Ampère–Maxwell law (1*d*), and the principle of conservation of charge (2).

What makes this view questionable is the assumption that, for *every* region $\Omega$ of space there exists some period of time during which the e/m field in $\Omega$ vanishes. This hypothesis is arbitrary and is not dictated by the theory itself. (It is likely that no such region exists in the Universe!) Therefore, the argument that led from relations (3) and (4) to relations (5) is not convincing since it was based on an arbitrary and, in a sense, artificial initial condition: that the e/m field is zero at some time $t=0$ and before.

Let us assume for the sake of argument, however, that there exists a region $\Omega$ within which the e/m field is zero for $t<t_0$ and nonzero for $t>t_0$. The critical issue is what happens at $t=t_0$; specifically, whether the functions expressing the e/m field are *continuous* at that moment. If they indeed are, the field starts from zero and gradually increases to nonzero values; thus, the line of reasoning that led from (3) and (4) to (5) is acceptable. There are physical situations, however, in which the appearance of an e/m field is so abrupt that it may be considered *instantaneous*. (For instance, the moment we connect the ends of a metal wire to a battery, an electric field suddenly appears in the interior of the wire and a magnetic field appears in the exterior. An even more "dramatic" example is *pair production* in which a charged particle and the corresponding antiparticle are created simultaneously, thus an e/m field appears at that moment in the region.) In such cases the e/m field is *non-continuous* at $t=t_0$ and its time derivative is *not defined* at this instant. Therefore, the line of reasoning that leads from (3) and (4) to (5) again collapses.

Note, finally, a circular reasoning in Stratton's approach. It is assumed that, in a region $\Omega$ where no e/m field exists, the second of relations (5) is valid identically.

This means that the vanishing of the electric field in *Ω* automatically implies the absence of electric charge in that region. This fact, however, follows from Gauss' law (1*a*); thus it may not be used *a priori* as a tool for proving the law itself!

Regarding charge conservation, we mentioned earlier that Eq. (2) is derived from the two non-homogeneous Maxwell equations, namely, Gauss' law (1*a*) for the electric field, and the Ampère–Maxwell law (1*d*). This means that the principle of conservation of charge is a *necessary* condition in order for the Maxwell system to be self-consistent. This condition is *not sufficient*, however, in the sense that it cannot replace any one of the system equations. Indeed, by the Ampère–Maxwell law and the conservation of charge there follows the *time derivative* of Gauss' law for the electric field [Eq. (4)]; this, however, does not imply that Gauss' law itself is valid. Of course, the reverse is true: *because* Gauss' law is valid, the same is true for its time derivative.

Our view, therefore, is that the Maxwell equations form a system of four independent PDEs that express respective laws of Nature. Moreover, the self-consistency of this system imposes two *necessary* (but *not sufficient*) conditions that concern the conservation of charge and the wave behavior of the time-dependent e/m field. In the next section the problem is re-examined from the point of view of Bäcklund transformations.

## 2. A Bäcklund-transformation view of Maxwell's equations

In previous articles [8,9] we suggested that, mathematically speaking, the Maxwell equations in empty space may be viewed as a Bäcklund transformation (BT) relating the electric and the magnetic field to each other. Let us briefly summarize a few key points regarding this idea. To begin with, let us see the simplest, perhaps, example of a BT.

The *Cauchy-Riemann relations* of complex analysis,

$$u_x = v_y \quad (a) \qquad u_y = -v_x \quad (b) \tag{6}$$

(where subscripts denote partial derivatives with respect to the indicated variables) constitute a BT for the *Laplace equation*,

$$w_{xx} + w_{yy} = 0 \tag{7}$$

Let us explain this: Suppose we want to solve the system (6) for $u$, for a given choice of the function $v(x,y)$. To see if the PDEs (6*a*) and (6*b*) match for solution for $u$, we must compare them in some way. We thus differentiate (6*a*) with respect to $y$ and (6*b*) with respect to $x$, and equate the mixed derivatives of $u$. That is, we apply the *integrability condition* (or *consistency condition*) $(u_x)_y = (u_y)_x$ . In this way we eliminate the variable $u$ and we find a condition that must be obeyed by $v(x,y)$:

$$v_{xx} + v_{yy} = 0.$$

Similarly, by using the integrability condition $(v_x)_y = (v_y)_x$ to eliminate $v$ from the system (6), we find the necessary condition in order that this system be integrable for $v$, for a given function $u(x,y)$:

$$u_{xx} + u_{yy} = 0.$$

In conclusion, the integrability of system (6) with respect to either variable requires that the other variable satisfy the Laplace equation (7).

Let now $v_0(x,y)$ be a known solution of the Laplace equation (7). Substituting $v=v_0$ in the system (6), we can integrate this system with respect to $u$. It is not hard to show (by eliminating $v_0$ from the system) that the solution $u$ will also satisfy the Laplace equation. As an example, by choosing the solution $v_0(x,y)=xy$ of (7), we find a new solution $u(x,y)=(x^2-y^2)/2+C$.

Generally speaking, a BT is a system of PDEs connecting two functions (say, $u$ and $v$) in such a way that the consistency of the system requires that $u$ and $v$ independently satisfy the respective, higher-order PDEs $F[u]=0$ and $G[v]=0$. Analytically, in order that the system be integrable for $u$, the function $v$ must be a solution of $G[v]=0$; conversely, in order that the system be integrable for $v$, the function $u$ must be a solution of $F[u]=0$. If $F$ and $G$ happen to be functionally identical, as in the example given above, the BT is said to be an *auto-Bäcklund* transformation (auto-BT).

Classically, BTs are useful tools for finding solutions of nonlinear PDEs. In [8,9], however, we suggested that BTs may also be useful for solving *linear systems* of PDEs. The prototype example that we used was the Maxwell equations in empty space:

$$\begin{aligned} &(a) \quad \vec{\nabla}\cdot\vec{E}=0 \qquad &&(c) \quad \vec{\nabla}\times\vec{E}=-\frac{\partial\vec{B}}{\partial t} \\ &(b) \quad \vec{\nabla}\cdot\vec{B}=0 \qquad &&(d) \quad \vec{\nabla}\times\vec{B}=\varepsilon_0\mu_0\frac{\partial\vec{E}}{\partial t} \end{aligned} \tag{8}$$

Here we have a system of four PDEs for two vector fields that are functions of the spacetime coordinates $(x,y,z,t)$. We would like to find the integrability conditions necessary for self-consistency of the system (8). To this end, we try to uncouple the system to find separate second-order PDEs for $\vec{E}$ and $\vec{B}$, the PDE for each field being a necessary condition in order that the system (8) be integrable for the other field. This uncoupling, which eliminates either field (electric or magnetic) in favor of the other, is achieved by properly differentiating the system equations and by using suitable vector identities, in a manner similar in spirit to that which took us from the first-order Cauchy-Riemann system (6) to the separate second-order Laplace equations (7) for $u$ and $v$.

As discussed in [8,9], the only nontrivial integrability conditions for the system (8) are those obtained by using the vector identities

$$\vec{\nabla}\times(\vec{\nabla}\times\vec{E})=\vec{\nabla}(\vec{\nabla}\cdot\vec{E})-\nabla^2\vec{E} \tag{9}$$

$$\vec{\nabla}\times(\vec{\nabla}\times\vec{B})=\vec{\nabla}(\vec{\nabla}\cdot\vec{B})-\nabla^2\vec{B} \tag{10}$$

By these we obtain separate wave equations for the electric and the magnetic field:

$$\nabla^2\vec{E}-\varepsilon_0\mu_0\frac{\partial^2\vec{E}}{\partial t^2}=0 \tag{11}$$

$$\nabla^2 \vec{B} - \varepsilon_0 \mu_0 \frac{\partial^2 \vec{B}}{\partial t^2} = 0 \qquad (12)$$

We conclude that the Maxwell system (8) in empty space is a BT relating the e/m wave equations for the electric and the magnetic field, in the sense that the wave equation for each field is an integrability condition for solution of the system in terms of the other field.

The case of the full Maxwell equations (1) is more complex due to the presence of the source terms $\rho, \vec{J}$ in the non-homogeneous equations (1*a*) and (1*d*). As it turns out, the self-consistency of the BT imposes restrictions on the terms of non-homogeneity as well as on the fields themselves. Before we get to this, however, let us see a simpler "toy" example that generalizes that of the Cauchy-Riemann relations.

Consider the following non-homogeneous linear system of PDEs for the functions $u$ and $v$ of the variables $x, y, z, t$:

$$\begin{aligned} & u_x = v_y \quad (a) && u_z = v_z + p(x,y,z,t) \quad (c) \\ & u_y = -v_x \quad (b) && u_t = v_t + q(x,y,z,t) \quad (d) \end{aligned} \qquad (13)$$

where $p$ and $q$ are assumed to be given functions. The necessary consistency conditions for this system are found by cross-differentiation of the system equations with respect to the variables $x, y, z, t$. In particular, by cross-differentiating ($a$) and ($b$) with respect to $x$ and $y$ we find that $u_{xx}+u_{yy}=0$ and $v_{xx}+v_{yy}=0$; hence both $u$ and $v$ must satisfy the Laplace equation (7). On the other hand, cross-differentiation of ($c$) and ($d$) with respect to $z$ and $t$ eliminates the fundamental variables $u$ and $v$, yielding a necessary condition for the terms of non-homogeneity, $p$ and $q$; that is, $p_t - q_z = 0$. This means that the functions $p$ and $q$ cannot be chosen arbitrarily from the outset but must conform to this latter condition in order for the system (13) to have a solution.

As an application, let us take $v=xy+zt$ (which satisfies the Laplace equation $v_{xx}+v_{yy}=0$) and let us choose $p=2t$ and $q=2z$ (so that $p_t - q_z = 0$). It is not hard to show that the solution of the system (13) for $u$ is then given by

$$u(x,y,z,t) = (x^2 - y^2)/2 + 3zt + C\,.$$

Notice that $u_{xx}+u_{yy}=0$, as expected.

Let us now return to the full Maxwell equations (1), which we now view as a BT relating the electric and the magnetic field and containing additional terms in which only the sources appear. As can be checked, there are now three nontrivial integrability conditions, namely, those found by applying the vector identities (9) and (10), as well as the identity

$$\vec{\nabla} \cdot \left( \vec{\nabla} \times \vec{B} \right) = 0 \qquad (14)$$

(the corresponding one for $\vec{E}$ is trivially satisfied in view of the Maxwell system). By (9) and (10) we get the non-homogeneous wave equations

$$\nabla^2 \vec{E} - \varepsilon_0 \mu_0 \frac{\partial^2 \vec{E}}{\partial t^2} = \frac{1}{\varepsilon_0} \vec{\nabla} \rho + \mu_0 \frac{\partial \vec{J}}{\partial t} \tag{15}$$

$$\nabla^2 \vec{B} - \varepsilon_0 \mu_0 \frac{\partial^2 \vec{B}}{\partial t^2} = -\mu_0 \vec{\nabla} \times \vec{J} \tag{16}$$

Additionally, the integrability condition (14) yields the equation of continuity (2),

$$\vec{\nabla} \cdot \vec{J} + \frac{\partial \rho}{\partial t} = 0 \tag{17}$$

expressing conservation of charge. Notice that, unlike (15) and (16), the condition (17) places *a priori* restrictions *on the sources* rather than on the fields themselves!

In any case, the three relations (15) – (17) are *necessary* conditions imposed by the requirement of self-consistency of the BT (1). As explained in Sec. 1, however, these conditions are *not sufficient*, in the sense that none of them may replace any equation in the system (1). In particular, the equation of continuity (17) may not be regarded as more fundamental than the Gauss law (1*a*) for the electric field.

## 3. Conclusions

Let us summarize our main conclusions:

1. The Maxwell equations (1) express four separate laws of Nature. These equations are mathematically consistent with one another but constitute a set of independent vector relations, in the sense that no single equation may be deduced by the remaining three. In particular, the physical arguments that attempt to render the two Gauss' laws "redundant" are seen to be artificial and unrealistic.

2. We consider the Maxwell equations as physically acceptable simply because the system (1) and all conclusions mathematically drawn from it represent experimentally verifiable situations in Nature. Among these conclusions are the conservation of charge and the conservation of energy (Poynting's theorem). It should be kept in mind, however, that conservation laws appear as *consequences* of the fundamental equations of a theory, and not vice versa. In particular, conservation of charge, in the form of the continuity equation (17), is a physically verifiable mathematical conclusion drawn from the Maxwell system (1) but it may not be regarded as more fundamental than any equation in the system. The same can be said with regard to the existence of e/m waves, expressed mathematically by Eqs. (11) and (12).

3. From a mathematical perspective, the Maxwell system (1) may be viewed as a Bäcklund transformation (BT) the integrability conditions of which (i.e., the *necessary* conditions for self-consistency of the system) yield separate (generally non-homogeneous) wave equations (15) and (16) for the electric and the magnetic field, respectively, as well as the equation of continuity (17). These integrability conditions are derived by *differentiating* the BT in different ways; hence they carry less information than the BT itself. Consequently, none of the integrability conditions may replace any equation in the Maxwell system.

## Acknowledgment

We thank Dr. Norman H. Redington of MIT for suggesting future directions for this research.

[1] https://arxiv.org/abs/1711.09969

[2] http://www.aemjournal.org/index.php/AEM/article/view/311

[3] http://nausivios.snd.edu.gr/docs/2016C.pdf